\documentclass[conference]{IEEEtran}

\usepackage{graphicx}
\usepackage[caption=false]{subfig}

\usepackage{algorithm}
\usepackage{algorithmicx}
\usepackage[noend]{algpseudocode}

\usepackage{graphics} 

\usepackage{enumitem}
\usepackage{cite}
\usepackage{amsmath,amssymb,amsfonts}
\usepackage{textcomp}
\usepackage{xcolor}
\def\BibTeX{{\rm B\kern-.05em{\sc i\kern-.025em b}\kern-.08em
    T\kern-.1667em\lower.7ex\hbox{E}\kern-.125emX}}

\begin{document}

\title{Elastic Composition of Crowdsourced \\ IoT Energy Services\vspace{-0.4cm}}

\author{\IEEEauthorblockN{Abdallah Lakhdari\IEEEauthorrefmark{1},
Athman Bouguettaya\IEEEauthorrefmark{2}, Sajib Mistry\IEEEauthorrefmark{2}, Azadeh Ghari Neiat\IEEEauthorrefmark{2}, Basem Suleiman\IEEEauthorrefmark{2}}
\IEEEauthorblockA{School of computer science,
University of Sydney\\
Sydney, Australia \\
Email: \IEEEauthorrefmark{1}alak8451@uni.sydney.edu.au,
\IEEEauthorrefmark{2}\{athman.bouguettaya, sajib.mistry, azadehgharineiat, basem.suleiman\}@sydney.edu.au}
\vspace{-1.0cm}
}



\maketitle

\begin{abstract}
We propose a novel type of service composition, called {\em elastic} composition which provides a {\em reliable} framework in a highly {\em fluctuating}  IoT energy provisioning settings. We rely on {\em crowdsourcing} IoT energy (e.g., {\em wearables}) to provide {\em wireless} energy to nearby devices. We introduce the concepts of {\em soft deadline} and {\em hard deadline} as key criteria to cater for an elastic composition framework. We conduct a set of experiments on real-world datasets to assess the efficiency of the proposed approach.
\end{abstract}
\vspace{-0.2cm}
\vspace{-0.45cm}
\section{Introduction}
The concept of \textit{Internet of Things} (IoT) has emerged as a result of the advancement of multiple technologies such as wireless communication, low-power sensors, and embedded systems \cite{atzori2010internet}. In IoT environment,  devices (aka, \textit{things}) capabilities are augmented with sensors, computing resources, and network connectivity \cite{cabrera2018services}. The service paradigm is congruent with the concept of IoT \cite{cabrera2017implementing}. In this regard, we leverage the service paradigm to abstract the \textit{functional} and \textit{non-functional} properties, i.e., Quality of Service (QoS) of IoT devices capabilities as {\em IoT services} \cite{bouguettaya2017service}.



The proliferation of IoT services has the potential to develop a self-sustained crowdsourced IoT ecosystem \cite{cabrera2018services}. People may exchange IoT services such as computing offloading, communication proxies, {\em energy sharing}, etc. These services are key building blocks to develop novel peer-to-peer sharing applications in a smart city.  For example, a set of co-located smartphones in a coffee shop may provide computing power or storage space for any resource-constrained IoT user \cite{ahabak2015femto}. A smartphone with depleting battery may get energy from nearby wearables {\em wirelessly} \cite{UmetsuNAFS19}\cite{TranM0B19}. This crowdsourced IoT ecosystem is similar to Uber where people share their cars for ridesharing.  
 In this paper, we focus on IoT energy crowdsourcing.

\textit{Crowdsourcing energy as a service} is the process of delivering outsourced wireless energy from IoT service {\em providers} to IoT users (i.e., {\em consumers}) \cite{Previouswork11}. Energy services have the potential of creating a \textit{green} service exchange environment. Energy could be harvested using renewable energy sources. For example, a smart shoe may harvest energy from the physical activity of its user~\cite{UmetsuNAFS19}\cite{choi2017wearable}\cite{gorlatova2014movers}. This energy may be used to charge nearby IoT devices \textit{wirelessly} \cite{he2013energy}. There exist several wireless energy transfer technologies which can deliver up to 3 Watts power within 5-meter distance to multiple receivers\footnote{https://www.energous.com/}\footnote{https://wi-charge.com/}. Wireless charging may provide a \textit{convenient} alternative as users do not need to look for power banks or sockets. 


{\em The composition} of IoT services is paramount in the crowdsourced IoT environment . A single crowdsourced IoT service may not fulfill the requirement of a consumer \cite{lakhdari2020composing}. {\em Multiple services need to be combined}. Composing crowdsourced IoT services requires the real-time selection and discovery of nearby IoT devices to accommodate an IoT service request \cite{Previouswork11}. Therefore, {\em space} and {\em time} attributes are crucial to model and select crowdsourced IoT services. {\em We focus on composing crowdsourced IoT  energy} services.   Existing service composition techniques in multiple domains such as Web services, cloud computing and mobile computing mainly focus on the deterministic properties of a service \cite{rao2004survey} \cite{deng2018composition}. However, they did not address issues arising from resource constrained IoT devices \cite{li2012modeling} \cite{abusafia2020incentive}. We identify the intrinsic characteristics of crowdsourced IoT energy services which are fundamentally different from traditional services as follows.  



\begin{itemize}[leftmargin=*]
    \item \textbf{Flexibility:} The temporal and spatial attributes of energy services and requests might be subject to different levels of flexibility \cite{brdiczka2010temporal}. An energy service provider might leave the confined area at any time. Consumers may stay longer than their predefined time interval to acquire more energy. 
    \item \textbf{Fluctuation:} IoT energy service providers may advertise their surplus energy  as an IoT energy service. However, they may use their IoT devices and provide energy at the same time. Energy services may fail due to the extreme usage of the IoT device by their owners. Providers might also cease provisioning services due to limited battery capacity \cite{li2012modeling}. 
    
    \item \textbf{Mobility:} Energy service providers and consumers must be available at the same time and location, i.e., within a predefined wireless transmission range, to ensure a successful wireless energy transfer. It is not always guaranteed to find nearby energy services which satisfy the required energy during a specified time interval. IoT energy providers may move and may not maintain the wireless transmission range with the consumer. Crowdsourced energy services may exhibit an intermittent provision behavior \cite{lakhdari2020fluid}.
\end{itemize}
Among above characteristics, fluctuation and mobility may impact the {\em reliability} of crowdsourced energy services \cite{lakhdari2020composing}\cite{abusafia2020Reliability}. The usage and the movement of IoT users may violate the advertised properties of  energy services. Therefore, we need to select the most reliable services to satisfy an energy request. 

We propose a \textit{composition} framework of crowdsourced IoT energy services to fulfill user's energy requirements within a specified time interval in confined areas e.g., coffee shop and restaurant. We focus on only {\em the fluctuation} of services and {\em the flexibility}  of consumers which may cause failures for the energy services.  In the future work, we will include the mobility of  IoT energy providers and consumers in a crowdsourced IoT environment. We propose an {\em elastic} composition of crowdsourced energy services considering the fluctuation of energy services and the flexibility of IoT users. The elastic composition algorithm extends our previous temporal knapsack-based composition algorithm \cite{Previouswork11}. Similar to 
traditional service composition techniques, the temporal knapsack-based composition algorithm relies on the initial description of services i.e., {\em advertisement} to generate the composition plans \cite{rao2004survey} \cite{deng2018composition}. In a crowdsourced IoT environment, these composition techniques may not be applied. In real-time, IoT energy services may differ from their advertisement due to the fluctuation of service providers behaviors.  A service provider may not be able to provide the advertised energy services due to various reasons, e.g., \textit{failures} and \textit{incorrect assessment} of available energy capacity \cite{zhao2016spatial}. Service composition plans need to be recomputed whenever a participating service fails. The real-time re-composition may require an {\em undetermined waiting time} for {\em available} IoT energy services which may extend the initial specified time interval of the energy request. It is not always guaranteed  to find nearby energy services during the time interval of the energy request to replace the failed services. Available services may be already reserved for other consumers. The elastic composition ensures the minimum waiting time by selecting and reserving  an optimal set of services among the fluctuating crowdsourced IoT energy services {\em ahead of time}. We consider the {\em reliability} to assess the performance of the services based on the fluctuating behavior of the device owners. The reliability assessment is used to select the optimal services.



The proposed composition framework leverages  the { \em flexibility} of IoT energy consumers to find a composition with the highest reliability. The charging time interval of consumers is defined based on their daily activity model for their convenience \cite{brdiczka2010temporal} \cite{banovic2016modeling} e.g., the time and the duration spent by IoT user in a coffee shop.  Usually, the pattern of time spent in regularly visited places e.g., coffee shops or food courts can be defined by a {\em range} between the { \em minimum} and the {\em maximum} time spent \cite{do2013places}. We incorporate the flexibility of IoT energy consumers based on the range of their usual stay in a confined area. We define two deadlines for the energy query time interval.  a) \textit{soft deadline}, and b) \textit{hard deadline}. The \textit{significance} of the elastic composition is to complete the IoT users' energy requirements in the dynamic crowdsourced IoT environment by preventing \textit{failure-prone} services within a flexible predefined time interval. If a satisfactory amount of energy could not be accumulated by composing reliable services during the soft deadline, the elastic composition may extend the composition interval up to the hard deadline. \textit{To the best of our knowledge, this is the first approach to design the elastic composition for crowdsourced IoT energy services}. The main contributions of this paper are:
\vspace{-0.15cm}
\begin{itemize}[leftmargin=*]

    \item A temporal composition algorithm of crowdsourced energy services in a predefined time interval.
    \item A multi-objective heuristic-based composition algorithm to maximize the reliability and minimize the time extensions between the soft and hard deadlines.
    
    \item Experimental evaluation and analysis are conducted based on real-world datasets to illustrate the effectiveness of the proposed elastic composition framework. 
    
\end{itemize}
\vspace{-0.23cm}
\subsection*{Motivation Scenario}\label{motivScen}
\vspace{-0.12cm}
\begin{figure*}
\centering
\includegraphics[width=0.7\textwidth]{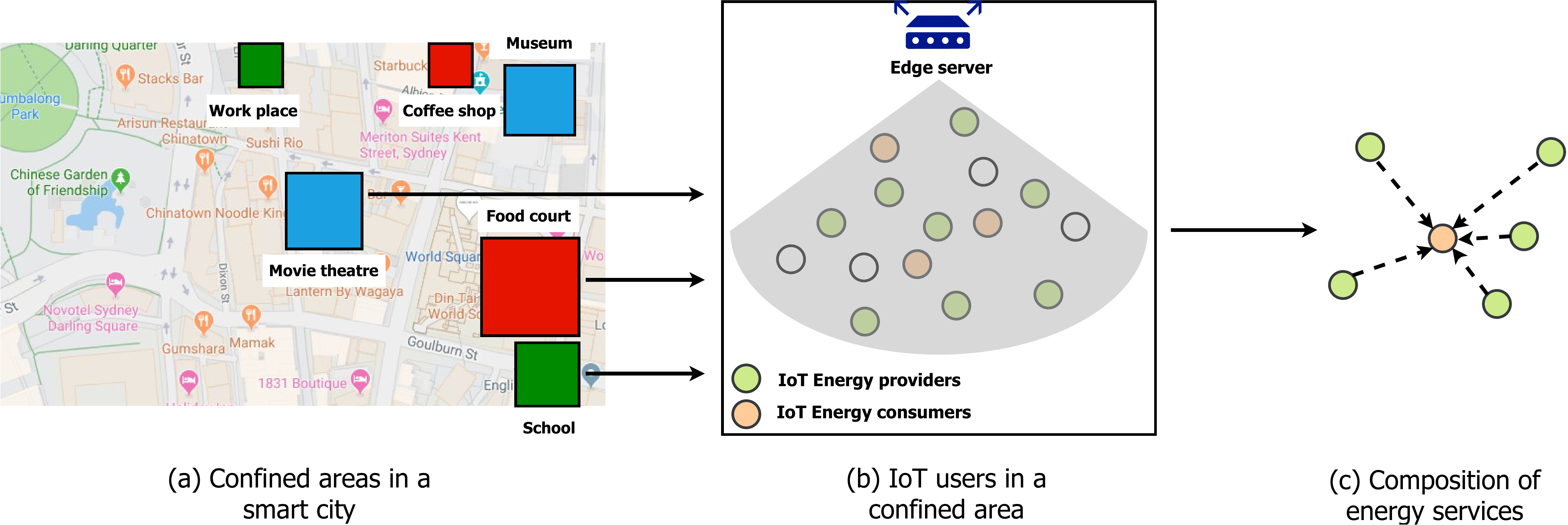}
\vspace{-0.1cm}
\caption{\small Composing crowdsourced energy services}
\vspace{-0.4cm}
\label{fig:deployment}
\end{figure*}
\begin{figure}
\centering
\includegraphics[width=0.45\textwidth]{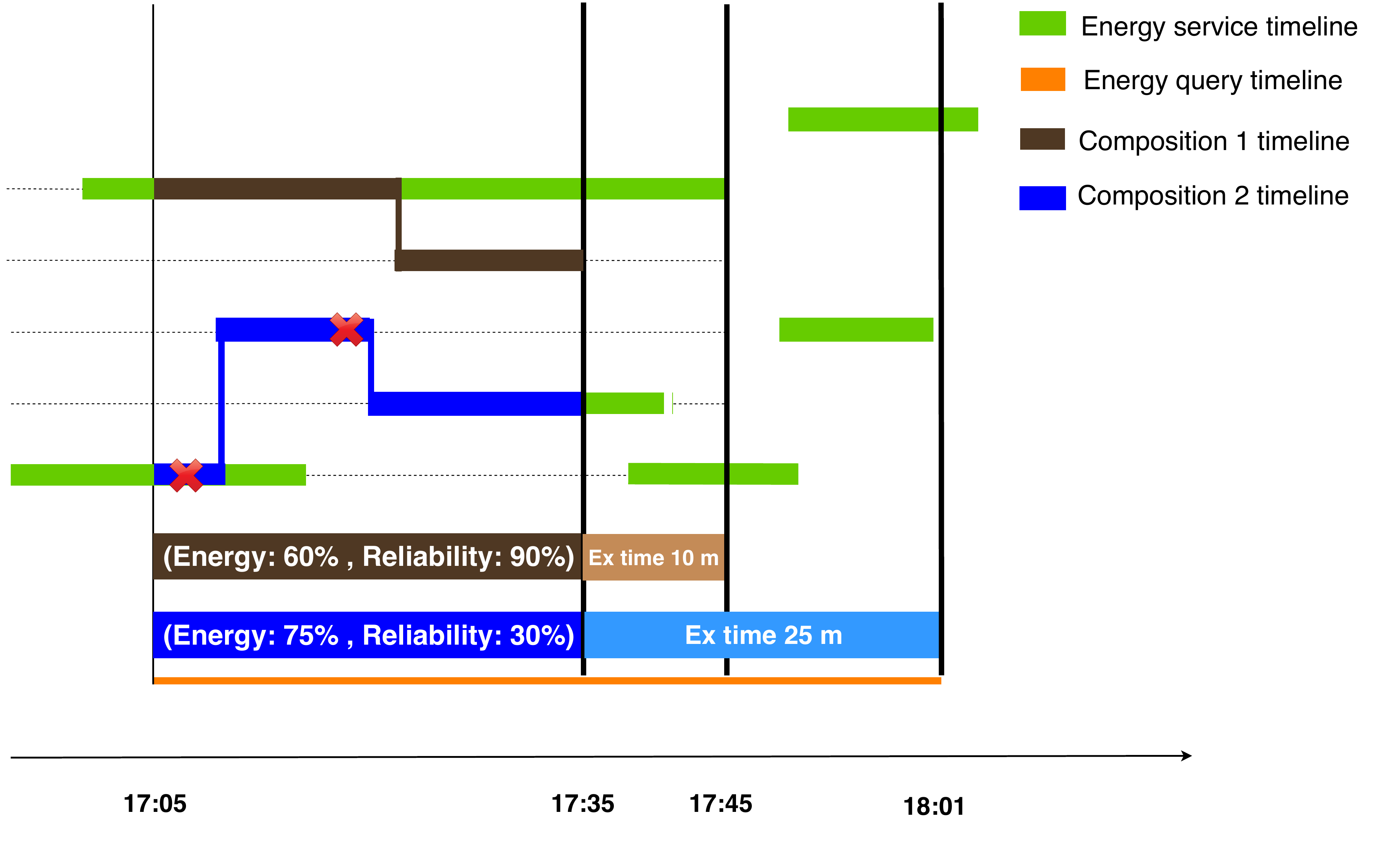}
\vspace{-0.4cm}
\caption{\small Different compositions with different scenarios}
\vspace{-0.5cm}
\label{fig:diff_compo}
\end{figure}

People may gather in different places (i.e., confined areas) in a smart city e.g., coffee shop, restaurant, work space, theatre, etc. (see Fig. \ref{fig:deployment} (a)). They may harvest energy by their {\em wearables} \cite{TranM0B19}. They may also share their spare energy wirelessly with nearby IoT devices. In a confined area, we assume that there are a set of energy service providers and consumers (see Fig. \ref{fig:deployment} (b)). Energy consumers may acquire energy from nearby IoT users. Let us consider a scenario of an IoT user in a coffee shop. A user may need to \textit{recharge} their smartphone. The IoT user casts a \textit{query} as \textit{User $x$ is looking for an amount of energy $E$  in the location $L$ during the time period $[t_s~,~t_e]$}.  The energy query is processed at the edge i.e., a router in a confined area (see Fig. \ref{fig:deployment} (b)). There exist five nearby energy service providers available in the same coffee shop who {\em advertised} their energy services (see Fig. \ref{fig:deployment} (c)).  The service advertisement includes various information about the provided service e.g., the service start time and end time, the provided energy amount. Fig. \ref{fig:diff_compo} illustrates the available services by their timelines. A neighbor service may \textit{drop out} at any moment due to multiple reasons e.g., a service provider may leave for an emergency, the provided energy amount may {\em fluctuate} because of the inordinate usage of the device owner. 

Multiple composition plans can be defined from the 5 available services. Let us assume that two different compositions have been provided $C_1$ in brown and $C_2$ in blue (see Fig. \ref{fig:diff_compo}). $C_1$ and $C_2$ provide $60\%$ and $75\%$ of the user’s energy requirement within the energy query duaration $[t_s~,~t_e]$ respectively. $C_1$ provides $60\%$ of the user’s energy requirement with $90\%$ of reliability. $C_2$ provides $75\%$ of the energy requirement with $30\%$ of reliability. $C_1$ composition needs {\em shorter} times extension to fulfill the required energy because of its service components which are less likely to fail according to the profiles of their providers. However, $C_2$ needs {\em longer} extra time to fulfill the energy requirement. Two participating services in $C_2$ are most likely to fail due to the excessive usage of the IoT devices by their owners. If these two services in $C_2$ fail, the expected energy from the composition $C_2$ drops drastically, e.g., from $75\%$ of the requested amount to $30\%$. \textit{It is challenging to recompose new services if a component service fail.} The previously advertised services for the initial query may not be available at the time of recomposition. The elastic composition assesses the fluctuation of crowdsourced IoT energy services in advance to select and reserve the less failure-prone services for the composition. Composing such services may provide a reliable composition which may require shorter time extension of the energy query time interval to accumulate the required energy. The elastic composition estimates the time extension of the query time interval based on failure likelihood of participating services. We reformulate our composition problem as \textit {"finding an optimal composition of crowdsourced IoT energy services which satisfies the energy requirement by maximizing the reliability and minimizing the query time extension".}







\vspace{-0.22cm}
\section{System model and problem formulation}\label{sysmdlpb}
\vspace{-0.1cm}
We present some preliminaries about crowdsourced energy services and queries. We then define the problem of elastic composition of crowdsourced IoT energy services.
\vspace{-0.18cm}
\subsection{Crowdsourced IoT Service Model}\label{sysmdl1}
\vspace{-0.08cm}
\paragraph*{Definition 1}
\textit{A crowdsourced  IoT energy service $ CES $} is a tuple $< Eid, Eownerid, F , Q >$: $Eid$ is a unique service ID, $Eownerid$ is a unique ID for the owner of the IoT device, $F$ is the set of $CES$ functionalities offered by an IoT device $D$. $Q$ is a tuple of $<q_1, q_2, ..., q_n>$ where each $q_i$ denotes a QoS property of $CES$ \cite{Previouswork11}.

\paragraph*{Definition 2}
\textit{Crowdsourced IoT energy  Quality of Service (QoS) Attributes} allow users to distinguish among crowdsourced IoT energy services. QoS parameters are defined as a tuple $< l, St,Et, DEC,I, Tsr, Rel_i>$: $l$ is the location of the consumer. $St$ and $Et$ represent the start time and end time of a crowdsourced IoT energy service respectively. $DEC$ is the deliverable energy capacity. $I$ is the intensity of the wirelessly transferred current. $Tsr$ represents the transmission success rate. $Rel_i$ represents the reliability QoS \cite{Previouswork11}. 

The spatio-temporal features of the IoT energy services (i.e, $l, St $ and $Et$) are defined based on the pattern of time spent in regularly visited places e.g., coffee shops or food courts  using their daily activity model in a smart city \cite{do2013places}. $DEC$ and $Rel_i$ are estimated based on the energy usage model of the IoT device \cite{zhang2010accurate}.  $I$ and $ Tsr$ are defined based on the specifications of the IoT devices providing services.


\textit{Definition 3: Crowdsourced IoT Energy Service Consumer Query}
is defined as a tuple $Q=< t_s, l, RE, CI, du, Dl_h>$. $t_s$ refers to the timestamp when the query is launched. $ l $ refers to the location of the energy service consumer. We assume that a consumer's location is fixed  after launching the query. $RE$ represents the required amount of energy. We also assume that the required energy is estimated based on an energy consumption model of the IoT device \cite{zhang2010accurate}. $CI$ is the maximum intensity of the wireless current that a consuming device can receive.  $du$ refers to a user-defined charging period. The initial charging period is considered as a soft-deadline $Dl_s$. $Dl_h$ represents the hard-deadline. It is the longest period of time, an energy consumer may wait for charging. Similarly, We define these two deadlines based on the patterns captured by the daily activity model \cite{do2013places}. Assuming that the energy query comes from energy poor IoT devices, our proposed  framework is run at the closest edge node to provide the optimal composition of nearby crowdsourced IoT energy services. Running the elastic composition framework at the edge significantly reduces the latency and the required energy for the composition.      



\vspace{-0.25cm}
\subsection{Crowdsourced IoT energy service composition problem}
\vspace{-0.08cm}

Given a set of crowdsourced IoT energy services $ S_{CES}=\{ CES_{1}, CES_{2}, \dots CES_{n} \} $ and a query $Q=< t_s, l, RE, CI, du, Dl_h>$ in a confined area, the problem is formulated as finding the most reliable composition of nearby energy services $ CES_{i} \in S_{CES}$ that can transfer the required amount of energy $Q.RE$ with the shortest extension of the predefined query duration $Q.du$ before the hard deadline $Dl_h$.
The available crowdsourced IoT energy services within the query duration may not provide a composition which can cater to the energy requirement during the query duration. In such a case, a query extension needs to be considered. The query might be extended to find any energy services that are very close to the specified soft deadline $Dl_s$ (i.e., the end of the initial query duration $Q.du$). 

\vspace{-0.2cm}
\section{Elastic composition of IoT Energy Services}
\begin{figure}
\centering
\includegraphics[width=0.45\textwidth]{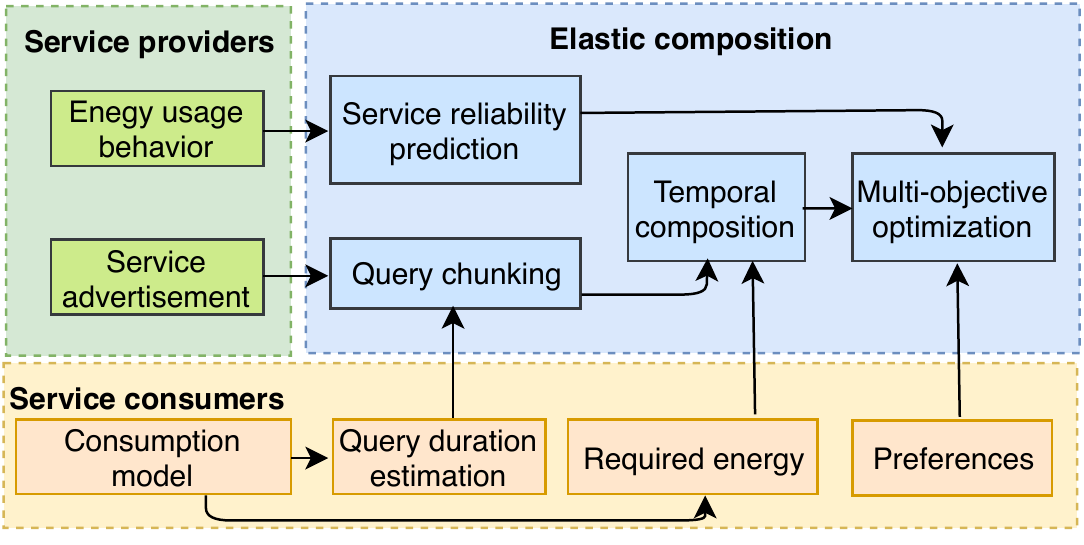}
\vspace{-0.1cm}
\caption{\small Elastic composition of crowdsourced IoT energy services}
\vspace{-0.5cm}
\label{fig:frameworkk}
\end{figure}

\vspace{-0.1cm}
Fig. \ref{fig:frameworkk} presents the building blocks of our proposed framework. The framework consists of three components: \textit{Service providers}, \textit{Service consumers}, and \textit{Elastic composition}. The elastic composition component takes as input, the advertised services and the energy query to perform the spatio-temporal composition. The composition considers the reliability of services and the flexibility of the charging waiting time of consumers. The \textbf{temporal composition} module provides all the possible compositions of the \textbf{advertised energy services} according to the temporal knapsack based composition algorithm \cite{Previouswork11}. The query duration is \textbf{chunked} according to the temporal features of the advertised services and the \textbf{query duration estimation}. We provide detailed explanation further in section \ref{chunkingggg}. Service providers use the energy service model to advertise their wireless energy services. Energy services are profiled based on the \textbf{usage behavior} of the IoT device owners to evaluate their \textbf{reliability}. We assume that energy consumers define their \textbf{energy requirements} and their charging waiting time (i.e., query duration) based on predefined \textbf{energy consumption models} \cite{zhang2010accurate}\cite{peltonen2015energy}. The \textbf{preferences} module of service consumers reflects their flexibility in time which is represented by the soft deadline $Dl_s$ and hard deadline $Dl_h$. The \textbf{multi-objective optimization} module selects the most reliable temporal compositions. In the following, we explain the concept reliability of crowdsourced IoT energy services and the building blocks of the elastic composition framework. 

\vspace{-0.2cm}
\subsection{Reliability of crowdsourced IoT energy services}\label{profiling}
\vspace{-0.1cm}

The advertised energy services might be unreliable because of the fluctuating behavior of the IoT devices owners. IoT devices are used by their owners during service provisioning. A service provider may not provide its \textit{advertised} energy service due to the excessive usage of its IoT device. The reliability refers to the probability that an IoT energy service will be successfully delivered with the same advertised QoS parameters. We define reliability probability of a crowdsourced IoT energy service based on the providers' history of usage and provision of their IoT resources \cite{ li2012modeling}. The energy capacity of IoT devices changes over time. It also may be different from an IoT user to another \cite{carroll2010analysis} \cite{peltonen2015energy}.

We rely on the regularity of energy usage to measure the reliability of energy services. We define $EUB_i $ as the energy usage behavior parameter of a device owner $i$. The values of $EUB_i$ range between 0 and 1. They can be calculated by several energy usage models which have been proposed to capture the energy usage regularity IoT devices \cite{oliver2011empirical}\cite{peltonen2015energy}. A number of  usage patterns is defined for IoT device. For example, Carroll et al.~\cite{carroll2010analysis} define different usage patterns of IoT devices. Providers usually follow one of these patterns as follows: \textit{Suspend} i.e., not using their devices,  \textit{Casual} i.e., using them casually with few functionalities, or \textit{Regular} i.e., using them with a predictable usage behavior. 

The provision history of a provider is also a good indicator of the service reliability. We consider the advertised energy without any QoS fluctuation or failure. We define $PB_i$ as the provision success of the provider $i$ by the ratio: 
$PB_i = \frac{SS_i}{TPS_i}$
where $SS_i$ is the number of successfully provisioned services. $TPS_i$ is the total number of previously provisioned services. 

The reliability of an IoT energy service $i$ can be estimated by: $Rel_i = EUB_i ~.~PB_i$. where the  $ EUB_i $ and $ PB_i$ are the energy usage behavior parameter and the provision history of its provider $i$ respectively,

\vspace{-0.2cm}
\subsection{Elastic Composition framework}
\vspace{-0.12cm}
We use the reliability score of crowdsourced IoT energy services to estimate the failure likelihood of component services.  Wireless energy could be received by IoT devices in two modes. Single wireless current or multiple wireless currents at a time \cite{na2018energy}. In this work, we focus on the single reception. Our future work will focus on receiving wireless energy from multiple providers. Therefore, we assume that an energy consumer acquires energy wirelessly from only one provider at a time. To present our composition approach, (i) we start by demonstrating the temporal chunking of a query duration; (ii) we then present the reliability model to distinguish between all the possible composite services; and finally (iii) we describe our heuristic-based composition approach to select an optimal temporal composite service.
\subsubsection{\textbf{Temporal chunking of the query duration}}\label{chunkingggg}
\begin{figure}
\centering
\includegraphics[width=.30\textwidth]{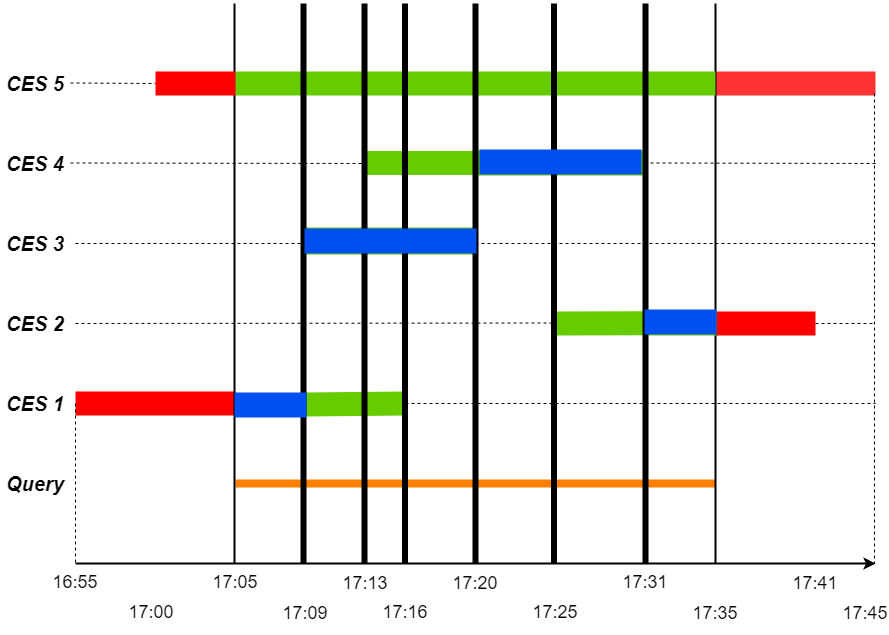}
\vspace{-0.1cm}
\caption{\small  Composing services}
\vspace{-0.5cm}
\label{fig:confidence}
\end{figure}


We first select all the available services between the start and the end time of an energy query $Q$. All the selected services are gathered in a set called $NearbyS$. We assume that IoT energy services are {\em on demand} services. They can be decomposed and consumed \textit{partially}, i.e., there is no lock-in contracts \cite{atzori2010internet}. A partial service $ps_i$ is a part of an  original service $s_i$ within a chunk. The energy consumer may switch to any other energy service within the query duration. To chunk the query duration, we follow our previous temporal knapsack algorithm \cite{Previouswork11}. We define all the possible {\em timestamps} where the consumer may need to switch to better services in terms of the provided energy. Each timestamp is either the start time or the end time of available services. We divide the query duration into several \textit{time chunks} based on these timestamps (see vertical lines in Fig. ~\ref{fig:confidence}). For example, the start time of $CES3$ defines the first chunk. The start time of $CES4$ defines the second chunk. An energy consumer may switch to a newly available service if a new service is better than the current service.
Some chunks may have a very short period of time. We define a minimum length $MinLch$ for a chunk considering the required time for the connection establishment between the consuming and providing devices.

\subsubsection{\textbf{Assessment of composite energy services}}\label{assessment}
The example in Fig. \ref{fig:confidence} illustrates 864 possible compositions for 5 available services during a 30 minutes query and 7 chunks $ch_i$  defined by the aforementioned chunking function. The total number of composite energy services $AllComp$ could be identified by multiplying the number of partial services at each chunk $ ParS_i$ for all the chunks $ AllComp = \prod_{i=1}^nParS_i$ where $n$ is the number of chunks. (e.g., $864=(2\in ch_1)$x$(3\in ch_2)$x$(4\in ch_3)$x$(3\in ch_4)$x$(2\in ch_5)$x$(3\in ch_6)$x$(2\in ch_7)$). Each candidate composite energy service $Comp_j$ is represented by a set of partial services defined by the query chunks \cite{Previouswork11}. We consider the QoS of services and the reliability of service providers (see definition 2 in section  \ref{sysmdlpb} and Section  \ref{profiling}) to distinguish between candidate composite energy services. We define the following rules to assess the aggregate reliability and the total provided energy of $Comp_j$:


\begin{itemize}[leftmargin=*]
    \item \textbf{Total energy capacity ($TEC$):} $TEC$ is the sum of the energy capacity of all the component partial services.
   \vspace{-0.2cm}
    \begin{equation}
    \footnotesize TEC(Comp_j) = \sum_{i=1}^{m} DEC(ps_i) 
     \vspace{-0.2cm}
    \end{equation}
    
    where $m$ is the number of component energy services. $ DEC(ps_i)$ is the deliverable energy capacity by $ps_i$, it is calculated based on the start time $st(ps_i)$ and end time of $et(ps_i)$, the energy intensity $I(ps_i)$, and the transmission success rate $Tsr(ps_i)$ (see Section \ref{sysmdl1})
    \begin{equation}
        \footnotesize DEC(ps_i) = (et(ps_i)-st(ps_i))~.~I(ps_i)~.~Tsr(ps_i)
    \end{equation}
    
    \item \textbf{Aggregate reliability score:}  Aggregate reliability score $AgR$ of $Comp_j$ is estimated based on the reliability of each component partial service. However, component partial services contribute differently to the total acquired energy amount and their allocated period of time within the query duration $Q.du$. $AgR$ of a composite energy service $Comp_j$ with $m$ component energy services is calculated by the following formula:
\end{itemize} 
    \begin{equation}
       \footnotesize AgR(Comp_j) = \frac{1}{m}\sum_{i=1}^{m} \frac{DEC(ps_i)}{TEC(Comp_j)}\times~\frac{du(ps_i)}{Q.du}\times~ Rel(ps_i)
    \end{equation}
    
    where $Rel(ps_i)$  is the reliability of the partial energy service $ps_i$. It is the same reliability of the initial service of $ps_i$. $du(ps_i)=et(ps_i)-st(ps_i)$ is the time period of the partial service. The fraction $\small{\frac{DEC(ps_i)}{TEC(Comp_j)}}$ is to quantify the contribution of $ps_i$ in terms of provided energy. The fraction $\small{\frac{(et(ps_i)-st(ps_i))}{Q.du}}$ is also used to quantify the participation of the allocated time for $ps_i$ from $Q.du$.         

We utilize the assessment rules to estimate the required time extension after the end of the query duration $Q.du$ for a particular composite energy service $Comp_j$. Intuitively, the required time extension relies on the remaining required energy amount to satisfy the energy query $Q.RE$. In our approach, we claim that the required amount of energy also relies on the aggregate reliability of composite energy services. A composite service $Comp_j$ with low reliability score means that the service is more prone to failures. If a component service fails, the acquired energy $TEC(Comp_j)$ will be affected. We estimate the remaining energy $RemRE$ as follows.
\begin{equation}
\footnotesize RemRE(Comp_j) = Q.RE- TEC(Comp_j).Agr(Comp_j)    
\end{equation}
We use the remaining required energy $RemRE(Comp_j)$ to reserve IoT energy services after the initial query duration $Q.du$. The query duration extension is estimated based on the reserved services. We consider the mean of the energy intensity $E(I)$ and the mean of the transmission success rate $E(Tsr)$ of all available services within the period of time $s_i \in [Q.t_s+Q.du~,~Q.t_s+2Q.du]$. We consider $2Q.du$ as the maximum tolerated query extension. 
\begin{equation}
    \footnotesize ExtQ(Comp_j) = \frac{RemRE(Comp_j)}{E(I(s_i))E(Tsr(s_i))}
\end{equation}

At the end, each candidate composite energy service $Comp_j \in AllComp$ is described by the total energy capacity $TEC(Comp_j)$  provided by the end of the query duration $Q.du$ with a certain aggregate reliability $AgR(Comp_j)$ and the estimated required time extension $ExtQ(Comp_j)$.

\subsubsection{\textbf{Heuristic based elastic composition}}

The proposed composition framework aims to select composite service $Comp_j \in AllComp$ which:
\begin{itemize}[leftmargin=*]
    \item maximize $AgR(Comp_j)$, and minimize $ExtQ(Comp_j)$

\item Subject to: 
\begin{enumerate}
    \item \small{$TEC(Comp_j) \geq Q.RE$}, \small{$Comp_j.st \geq Q.t$,}
    \item \small{$ExtQ(Comp_j)\leq Dl_h$.}
\end{enumerate}
\end{itemize}
The total number of candidate composite energy services $|AllComp|$ might be exponential even for a small number of services (see Section \ref{assessment}). Our temporal composition problem is reformulated as a multi-objective optimization problem. We present a heuristic to estimate the \textit{Pareto front set $PF$} represented by the optimal candidate composite services 
$AllComp$ is considered as the solution search space and each candidate composite service $Comp_j \in AllComp$ is defined with a set of attributes $Att$. We consider the attributes $att_1 = ExtQ$ and $att_2 = AgR$. We define the \textit{Pareto front set $PF$} using \textit{dominant} candidate services. 
\paragraph*{Definition 4} \textit{Dominant candidate composite energy service.} A composite energy service $Comp_i$ dominates another composite energy service $Comp_j$, denoted as $Comp_i > Comp_j$, if $Comp_i$ is as good or better than $Comp_j$  for all attributes i.e., $\forall ~att \in Att : Comp_i  \geq Comp_j $ and $\exists ~\small{att}\prime \in Att : Comp_i  > Comp_j$.

\paragraph*{Definition 5} \textit{Pareto front set $PF$.} is a subset of candidate composite services that are not dominated by any other composite service: 
\begin{equation}
 \footnotesize
 \begin{split}
     PF = \{ Comp \in AllComp 
 | \neg \exists \small{Comp}\prime \in AllComp :  ~~\\
 \small{Comp}\prime > Comp \}   
 \end{split}
\end{equation}

The proposed heuristic algorithm (see Algorithm \ref{heuristic} ) aims to reduce the solution search space of candidate composite energy services. The intuitive idea of our heuristic is (i) to reduce the number of chunks and  (ii) to consider a reduced number of partial services for each chunk. The remaining composite services are considered the most promising candidates $PromComp$. To illustrate this reduction, let us consider a scenario of a query with 6 chunks. Each chunk contains at least 4 partial services. If the number of chunks is reduced to 4 and only the top 3 partial services are considered for each chunk, the number of possible candidate solutions will decrease from to $4^6 =4096$  to $3^4=81$.
Algorithm \ref{heuristic}  merges chunks where the service providing maximum energy does not change over two consecutive chunks (see  blue segments in Fig. \ref{fig:confidence} and lines [6-10] in algorithm \ref{heuristic} ). Because most likely the same service is going to be selected in both chunks . In our heuristic-based optimization, partial services are ranked for each one of the new chunks (lines[14-17] in algorithm \ref{heuristic}). We define the Pareto front set of partial services in each chunk $TopK_i$. As a result, the number of candidate composite services is reduced significantly $|PromComp| << |AllComp|$ (line 21). We run an exhaustive search on the set $PromComp$ using \textit{ Definition 5} to find $PF$ set of the optimal candidate composite services. The following utility function $u$ is defined  to ponder the provided energy $DEC(Comp_i)$ and the reliability $Rel(Comp_i)$ of all Pareto front composite services based on the consumers' preferences.
\begin{equation}
  \footnotesize u(Comp_i)=w_e(TEC(Comp_i))+w_r(Rel(Comp_i))  
\label{eq:utility}
\end{equation}
where $w_t$ and $w_r$ represent the energy and reliability weights respectively. 
The proposed framework compromises the estimated energy capacity and the time extension and provides the optimal composition following three different preference strategies. Each strategy introduces preferences based on users' attitude towards failures. Consumers may be greedy and maximize energy regardless the reliability. They also may be risk-averse and consider reliability or neutral. 

\begin{algorithm}[h!]
	\caption{Heuristic-based space reduction}\label{heuristic}
	\begin{algorithmic}[1]
\footnotesize
\State \textbf{Input: }$ Q.l$, $Q.t_s$ , $Q.du$, $NearbyS$
		\State \textbf{Output: }$PromComp$
		\State \text{ // Chunking the query duration}
		\State $Chunks \gets Chunking(Q.l, Q.t , Q.d, NearbyS) $
		\State \text{ // Update chunking by maximum services}
		\State \textbf{For all} $C \in Chunks$  
		\If        {($ Max(C_i) = Max(C_{i+1}) $ ) }
		\State $C_i.et  \gets C_{i+1}.et$
		\State \text{Delete ($C_{i+1}$)}
		\EndIf
		\State \textbf{EndFor}
		\State \text{ // Updated chunks set is called \textbf{\textit{updatChunks}}  }
		\State \text{ // Set of top K $ps$ sets is called \textbf{\textit{New Space}}  }
		\State \text{$New~space \gets \emptyset  $}
		\State \textbf{For all} \text{$ updatC_i \in updatChunks $} \textbf{do}
		\State \text{$u$-based Sort($ All ps_j \in updatC_i $)}
		\State \text{$TopK_i \gets ps_j \in updatC_i $}
		\State \text{$New~space \gets New~space \cup TopK_i $}
		\State \textbf{EndFor}
		\State \text{ // The Cartesian product of \textbf{\textit{New Space}} n sets }
		\State \text{ // $X$ denotes the Cartesian product}
		\State \text{$PromComp = X_{i=1}^n New ~Space_i$}
		\State \textbf{End}

		
	\end{algorithmic}
\vspace{-0.1 cm}	
\end{algorithm}
\vspace{-0.3 cm}
\section{Experiments}
\vspace{-0.11cm}
\begin{table*}[t!]
\centering
\small
\begin{tabular}{|l|c|c|l|c|c|}
\hline
\multicolumn{3}{|c|}{\footnotesize{Crowdsourced IoT energy service}}                                                                                                          & \multicolumn{3}{c|}{\footnotesize{IoT energy query}}                                                                                                                                                                                  \\ \hline
\multicolumn{1}{|c|}{\footnotesize{QoS}} & \footnotesize{Dataset}                                                              & \footnotesize{value}                                                       & \multicolumn{1}{c|}{\begin{tabular}[c]{@{}c@{}}\footnotesize{Query parameters}\end{tabular}} & \footnotesize{Dataset}                                                              & \footnotesize{value}                                                       \\ \hline
\footnotesize{Start time}                & \footnotesize{Yelp}                                                                 & \footnotesize{Check-in}                                                    & \footnotesize{Start time}                                                                        & \footnotesize{Yelp}                                                                 & \footnotesize{Check-in}                                                    \\ \hline
\footnotesize{End time}                  & \footnotesize{Uniform distribution}                                                               & \footnotesize{Uniform distribution}                                                      & \footnotesize{Soft deadline}                                                                     & \begin{tabular}[c]{@{}c@{}}\footnotesize{Monte carlo}\end{tabular}              & \begin{tabular}[c]{@{}c@{}}\footnotesize{Monte carlo}\end{tabular}     \\ \hline
\footnotesize{Energy capacity}           & \begin{tabular}[c]{@{}c@{}}\footnotesize{Renewable energy sharing}\end{tabular} & \begin{tabular}[c]{@{}c@{}}\footnotesize{Provided energy}\end{tabular} & \footnotesize{Energy capacity}                                                                   & \begin{tabular}[c]{@{}c@{}}\footnotesize{Renewable energy sharing}\end{tabular} & \begin{tabular}[c]{@{}c@{}}\footnotesize{Provided  energy}\end{tabular} \\ \hline
\footnotesize{Reliability score}         & \footnotesize{Carat}                                                                & \footnotesize{Entropy}                                                      & \footnotesize{Hard deadline}                                                                     & \begin{tabular}[c]{@{}c@{}}\footnotesize{Monte carlo}\end{tabular}              & \begin{tabular}[c]{@{}c@{}}\footnotesize{Monte carlo}\end{tabular}     \\ \hline

\end{tabular}
\caption{Parameters of the experiments setting}
\vspace{-.6cm}
\label{tab:simparam}
\end{table*}

We evaluate the proposed composition techniques by an extensive set of experiments. First, we evaluate the {\em scalability} of each composition algorithm by measuring the computation time while varying the number of IoT energy services. Second, we evaluate the {\em efficiency} and the {\em effectiveness} of the proposed elastic composition technique by evaluating its performance in a failure-prone crowdsourced IoT environment. We compare the proposed approach with a brute-force, a greedy selection, and the temporal knapsack based composition \cite{Previouswork11}.  The greedy composition relies mainly on selecting and composing the best IoT energy services in terms of the user preferences e.g., maximum energy or shortest duration of a service composition. The temporal knapsack based composition algorithm chunks the initial query duration (see section \ref{chunkingggg}). The algorithm then selects the partial energy service providing the maximum energy capacity at each chunk \cite{Previouswork11}. 
\vspace{-0.25cm}
\subsection{Datasets and experiment environment}
\vspace{-0.1cm}
We evaluate the proposed composition techniques on real-world datasets. We create a scenario of crowdsourced IoT environment close to the reality to evaluate the performance of the elastic composition of IoT energy services. 

We use {\em Yelp}\footnote{https://www.yelp.com/dataset} dataset to define the {\em spatio-temporal features} of IoT energy services and queries by {\em check-in} and {\em check-out} timestamps of people to a confined area. The dataset contains people's check-ins information into different confined areas e.g., coffee shops, restaurants, libraries etc. We consider these people as IoT users who are either energy providers or consumers. For example, the start time $st$ of an energy service provided by an IoT user is the time of their check-ins into a coffee shop. Energy queries time $Q.t_s$ and duration $Q.du$ are also generated from check-in and check-out timestamps of customers. We use a {\em uniform distribution} to augment the check-in data and generate different {\em durations} for IoT energy services and queries. Usually, the time spent by people in regularly visited places e.g., coffee shops or food courts can be defined as a flexible time interval ranging from the { \em minimum} and the {\em maximum}. We use {\em Monte Carlo} simulation to generate soft and hard deadlines for the IoT energy queries.

To the best of our knowledge, there is no dataset about wireless energy sharing among human-centric IoT devices. We consider that crowdsourced IoT energy services are provided from wearables or the spare energy of the smartphone batteries of IoT users.  We create an IoT energy crowdsourcing environment close to reality based on a renewable energy sharing environment \footnote{https://data.gov.au/dataset}. A set of 25 houses daily harvest, consume and provide energy from their solar panels for two years [April 2012 to March 2014]. They are considered either as energy providers or consumers. Energy consumption and production is recorded every 30 minutes. Each house has 730 daily records ($365$x$2$). Each daily record has $48$x$2$ fields for the produced and the consumed energy for each day. The crowdsourced IoT energy service QoS parameters are defined based on these records. We normalize all the energy measurement values for all records from {\em Watt hour to miliampere hour (mAh)} to mimic the energy provided and consumed by IoT devices e.g., smartphone and wearables. The deliverable energy capacity $DEC$ of IoT energy services $S_i$ starting at $S_i.st$ and ending at $S_i.et$  is defined by {\em randomly} matching a daily record of a provider from the renewable energy sharing environment considering only the energy produced during the same period of time. Similarly, the energy requirement of a query $Q.RE$ is also generated and normalized from the daily energy consumption of the houses according to the query duration $Q.du$.

We use {\em Carat}\footnote{https://www.cs.helsinki.fi/group/carat/data-sharing/} dataset to simulate the fluctuating behavior of crowdsourced IoT energy services. Energy services are provided by IoT devices which are being used by their owners. Carat dataset contains time series of the state of charge (SoC) of 1000 smartphones. We use this dataset to estimate the reliability score of IoT energy services. The reliability score is calculated based on capturing the fluctuation of SoC time series using the entropy \cite{oliner2013carat}. We randomly map each IoT energy service with one of the 1000 obtained reliability scores. Table \ref{tab:simparam} and table \ref{tab:simStat} recapitulate the experiments parameters.
\begin{table}
\centering
\begin{tabular}{|l|l|}
\hline
\footnotesize{Parameter} & \footnotesize{Range of values} \\
\hline
\footnotesize{Confined areas}                & \footnotesize{8280}                                \\ \hline
\footnotesize{Queries}                       & \footnotesize{5000}                                \\ \hline
\footnotesize{Services}                      & \footnotesize{5000-50 000}                         \\ \hline
\footnotesize{Duration of a service}         & \footnotesize{10-60  minutes}                      \\ \hline
\footnotesize{Duration of a query}           & \footnotesize{5-120 minutes}                      \\ \hline
\footnotesize{Provided energy}               & \footnotesize{50-1000 mAh}                        \\ \hline
\footnotesize{Energy requirement}            & \footnotesize{100-800 mAh}                       \\ \hline
\end{tabular}
\caption{Statistics of the crowdsourced IoT environment}
\vspace{-.7cm}
\label{tab:simStat}
\end{table}


\vspace{-0.2cm}
\subsection{Scalability}
\vspace{-0.1cm}
\begin{figure*}[t!]
    \centering
    \subfloat{\includegraphics[width=.31\textwidth]{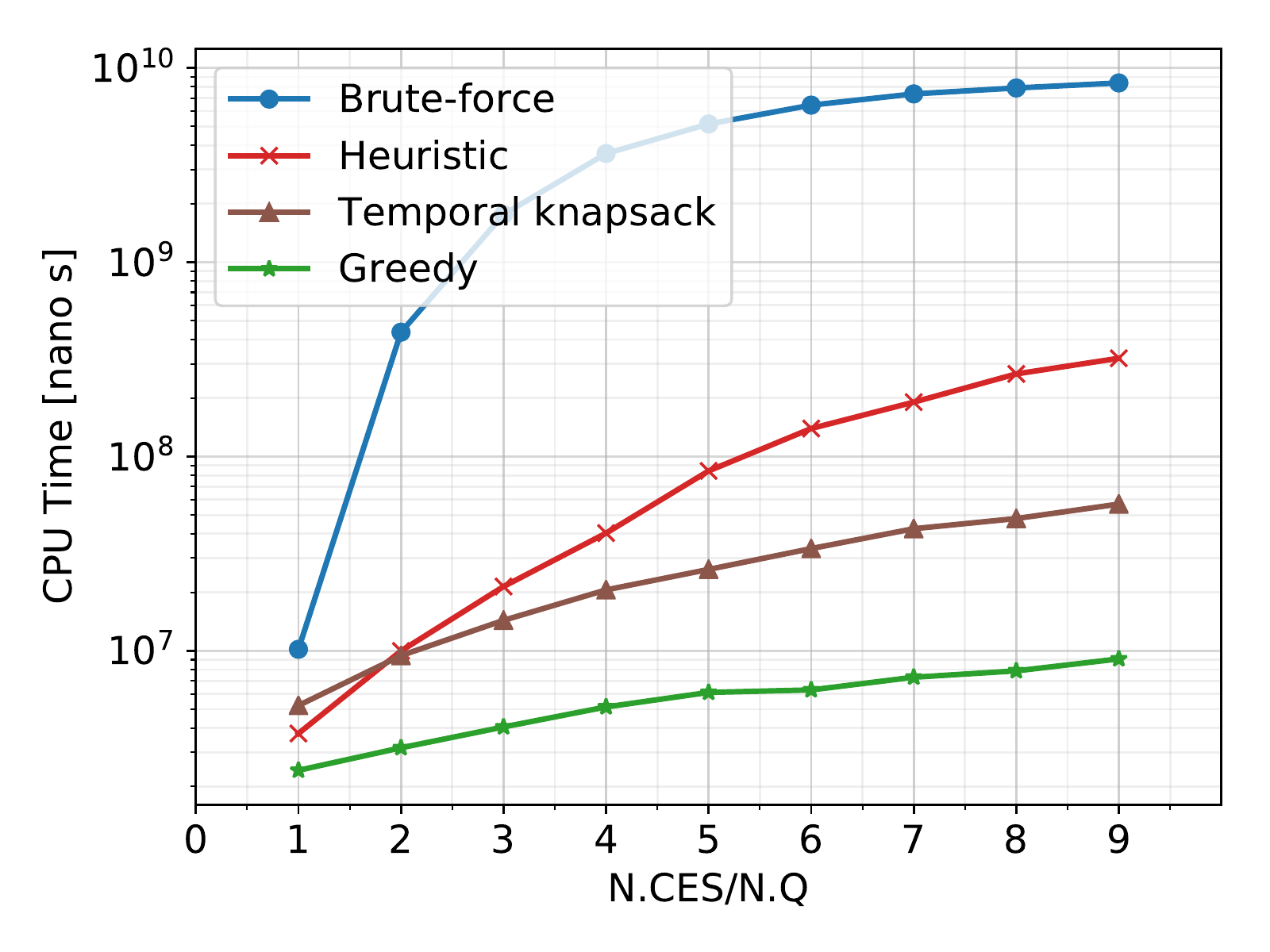}}
    \subfloat{\includegraphics[width=.31\textwidth]{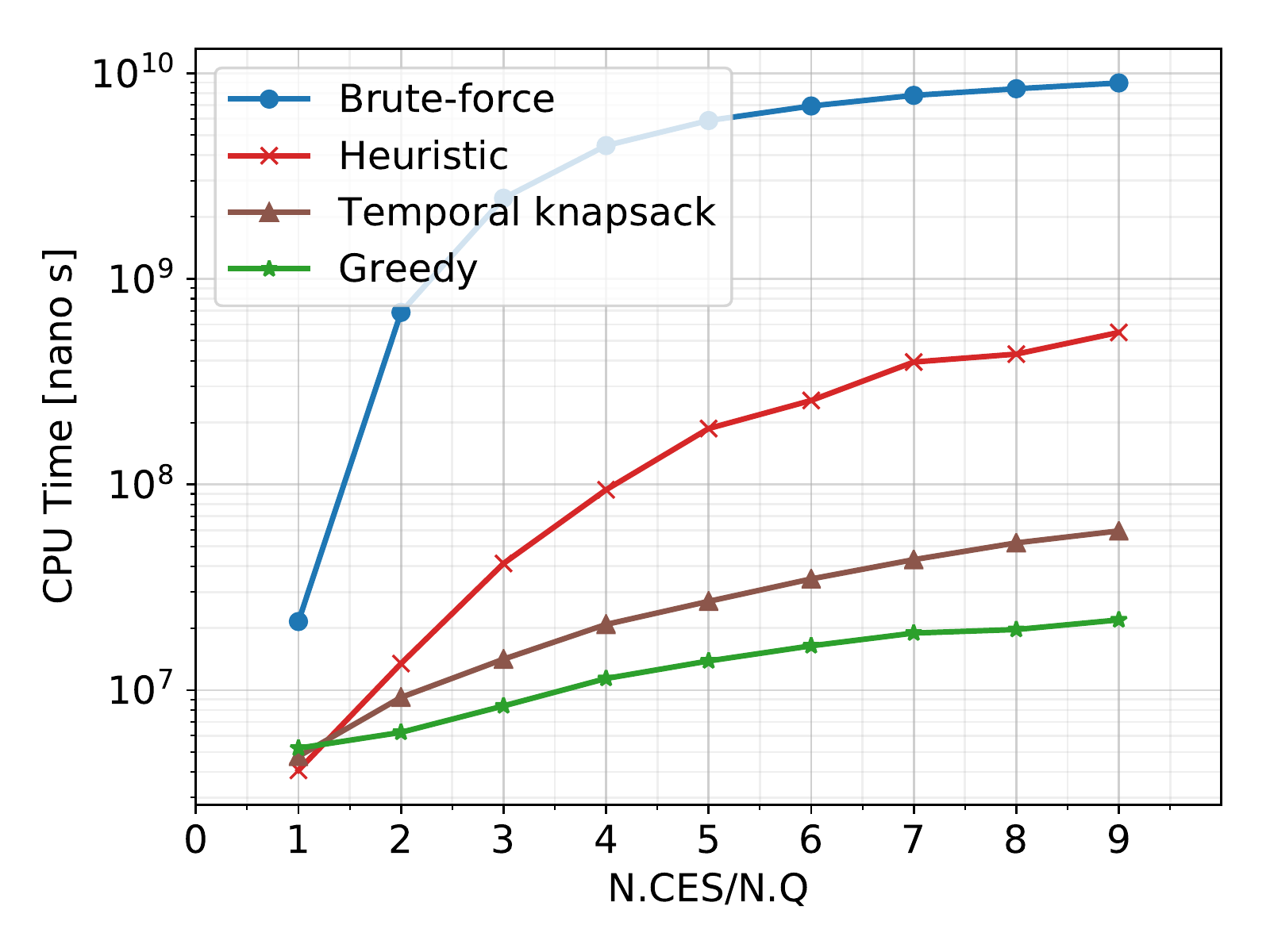}}
    \subfloat{\includegraphics[width=.31\textwidth]{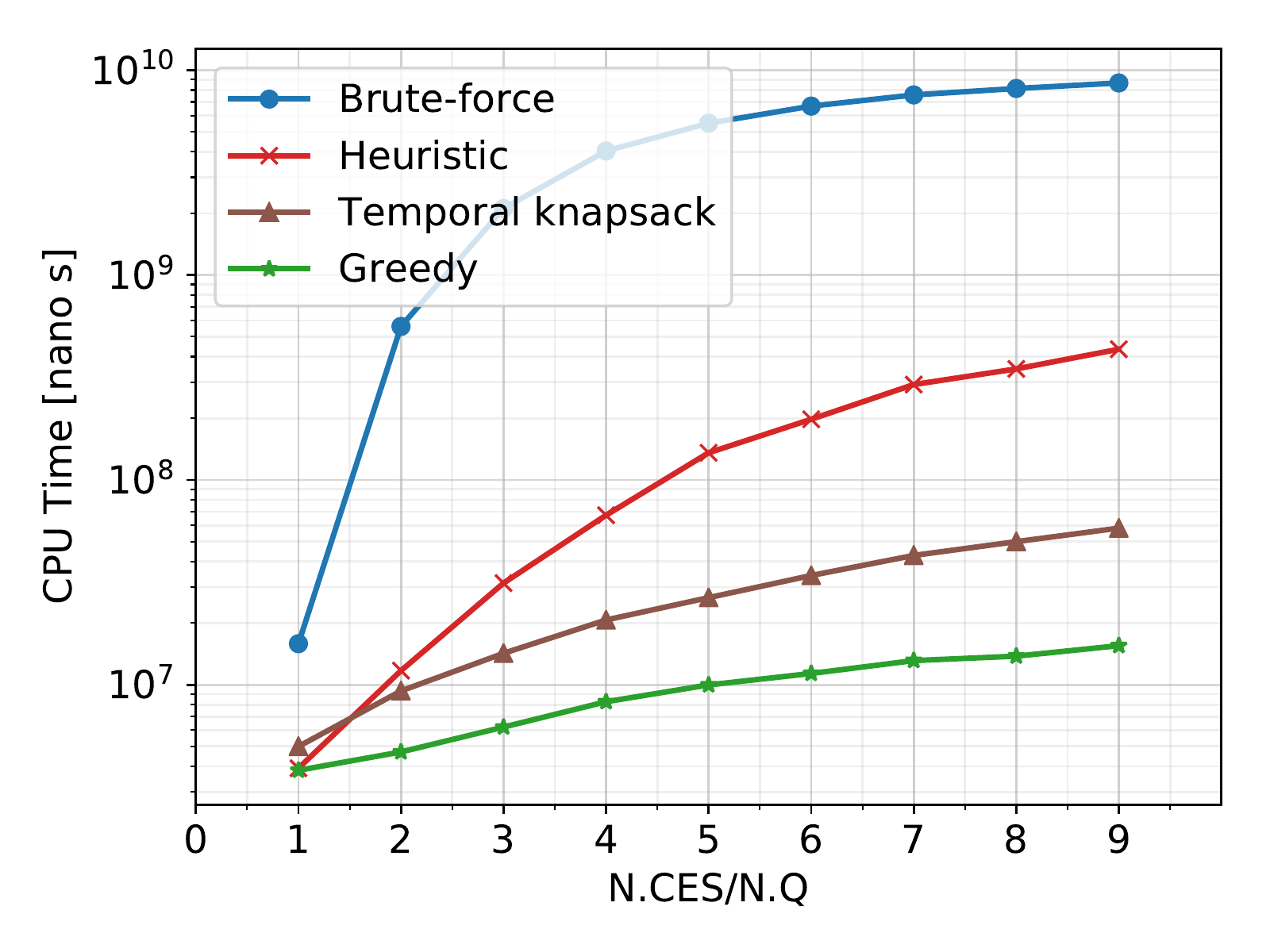}}
    \vspace{-0.2cm}
     \caption{\small CPU time for elastic composition (a) composing short services (b) composing long services (c) composing all services}
    \label{fig:cpu}
    \vspace{-0.4cm}
\end{figure*}
We investigate the computation time of the proposed composition approach by a large number of energy services. 50,000 different IoT users attending 8,000 confined area. The aim of the scalability evaluation is to ensure the {\em light weight implementation} of the elastic composition framework at edge servers (see Fig. \ref{fig:deployment} (b)) in section \ref{motivScen}. We fix the number of IoT energy queries and gradually increase the number of crowdsourced IoT energy services to observe the execution time behavior for the different composition algorithms. The total number of energy queries in all confined areas is 5000. We modify the ratio between the number of services and the number of queries  ($N.CES/N.Q$) among the IoT users from ${1}$ to ${9}$ which leads the total number of IoT energy services to vary from 5000 to 45,000 services. All evaluation algorithms are implemented in Java and conducted on a 3.60 GHZ Core i7 processor and 8 GB RAM under Windows 10.

 Fig. \ref{fig:cpu} shows the execution time of the four different composition algorithms, the brute-force and our proposed heuristic for the elastic composition, the previously proposed temporal fractional knapsack algorithm \cite{Previouswork11}, and a greedy composition algorithm. 
 We measure the average execution time of each composition technique to serve 5000 energy queries conducted in multiple scenarios, short services e.g. service duration between 10 to 30 minutes (see Fig. \ref{fig:cpu} (a)), long services e.g. service duration between 20 to 50 minutes (see Fig. \ref{fig:cpu} (b)), and all types of  services (see Fig.  \ref{fig:cpu} (c)).

These graphs highlight the light weight execution time of the heuristic based elastic composition which performs the temporal composition and the multi objective optimization. The increasing behavior of CPU time is expected for all composition algorithms because of the increase in the number of available services. First, we compare the heuristic based elastic composition with the greedy and the temporal knapsack-based composition algorithms. The heuristic algorithm is executed in a comparable time with the greedy and the temporal knapsack-based algorithm when the number of available services is relatively low ($N.CES/N.Q \in [1-3]$). However, the greedy and the temporal knapsack composition algorithms outperform the heuristic based composition when the number of available services is higher ($N.CES/N.Q \in [4-9]$). The greedy and the temporal knapsack algorithms \textit{do not consider the reliability} and they provide one single composition unlike our proposed {\em multi objective optimization heuristic  which considers the reliability and selects the most optimal composition of IoT energy services.} Second, the log scale representation accentuates the huge difference between CPU time of the  brute force and the heuristic-based elastic composition for all numbers of available services ($N.CES/N.Q \in [1-9]$). This difference is explained by the considerable {\em reduction of the candidate composite services space} before finding the Pareto front.


\vspace{-0.2cm}
\subsection{Efficiency}
\vspace{-0.1cm}
\begin{figure}
\centering
\vspace{-0.2cm}
\includegraphics[width=0.325\textwidth]{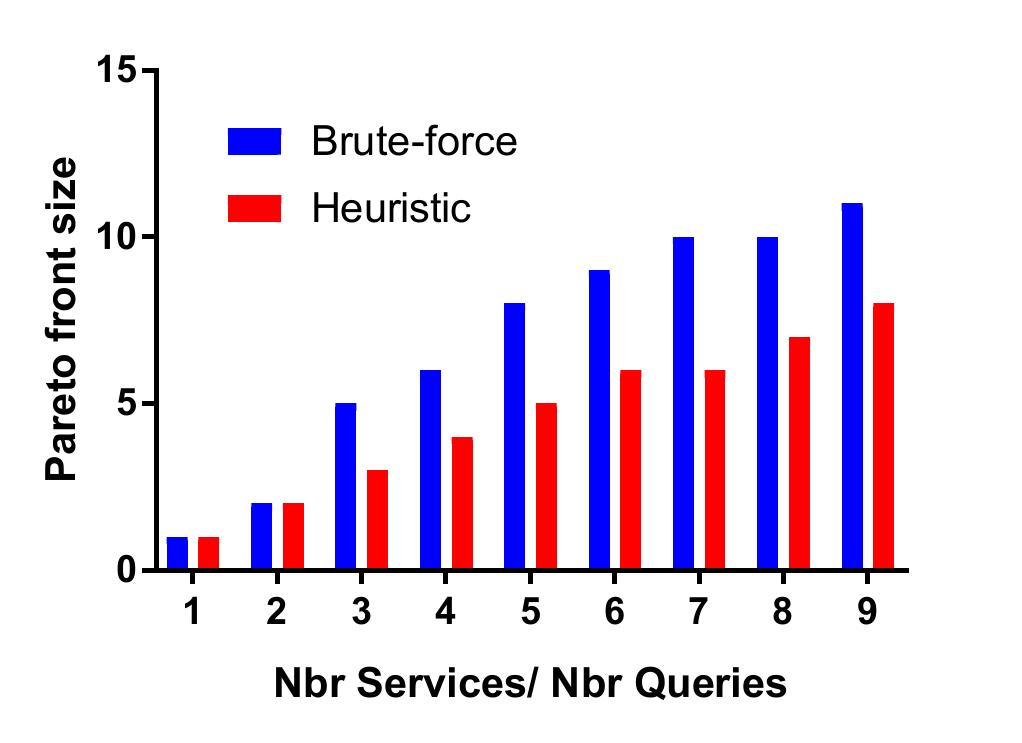}
\vspace{-0.2cm}
\caption{\small Number of candidate compositions in the Pareto front}
\label{fig:expected1}
\vspace{-0.4 cm}
\end{figure}

\begin{figure}
\centering
\includegraphics[width=0.31\textwidth]{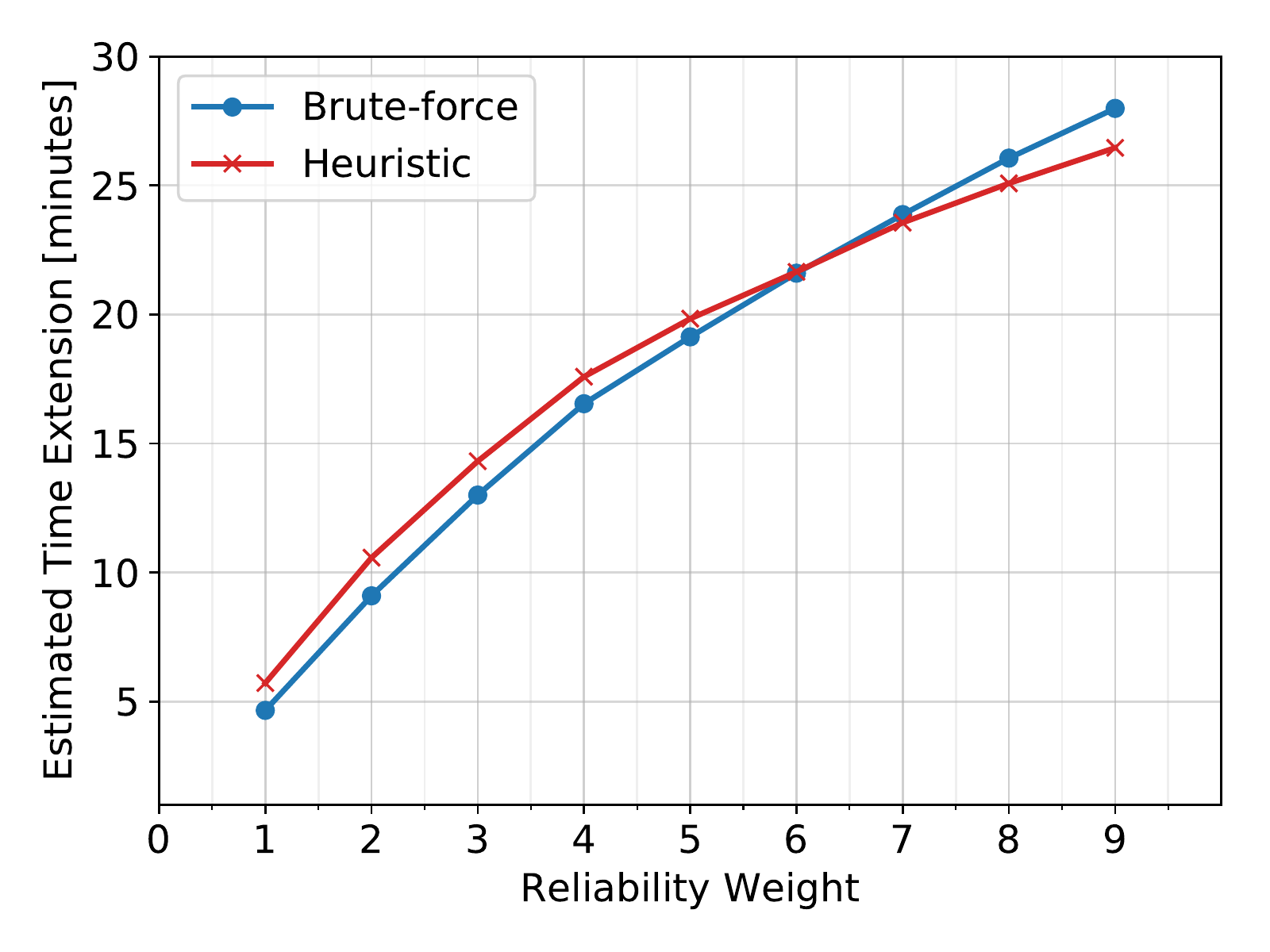}
\vspace{-0.2cm}
\caption{\small Query time extension estimation}
\label{fig:expected2}
\vspace{-0.6 cm}
\end{figure}



In this experiment, our ground truth is the brute force based elastic composition. It provides the optimal set of composite crowdsourced IoT energy services which fulfill an energy query. The optimal composite services are represented by their {\em reliability score} and the {\em expected time extension} of the query duration in a failure-prone environment of IoT energy services.  
\begin{figure*}[t!]
    \centering
    \subfloat{\centering\includegraphics[width=.31\textwidth]{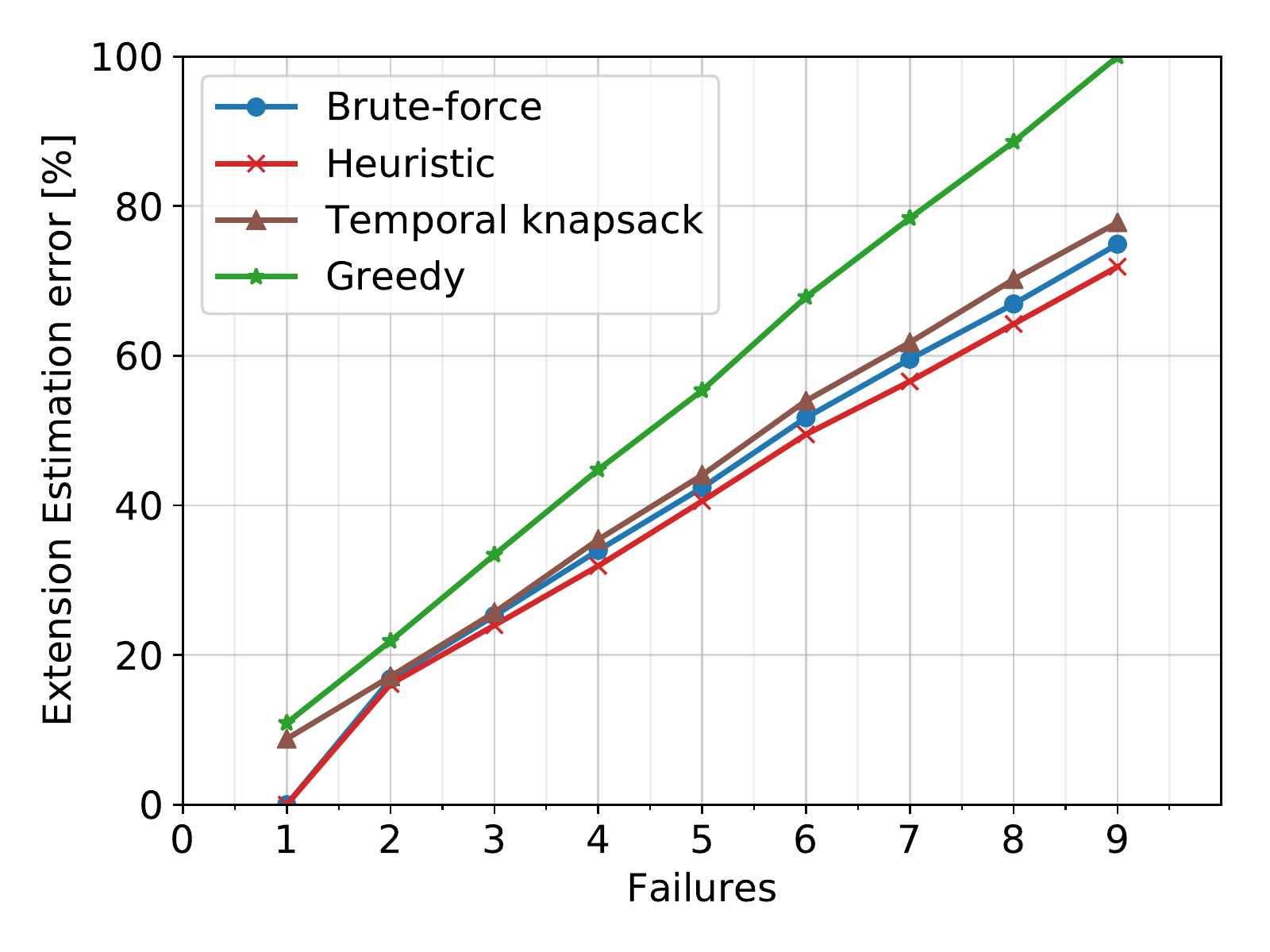}}
    \subfloat{\centering\includegraphics[width=.31\textwidth]{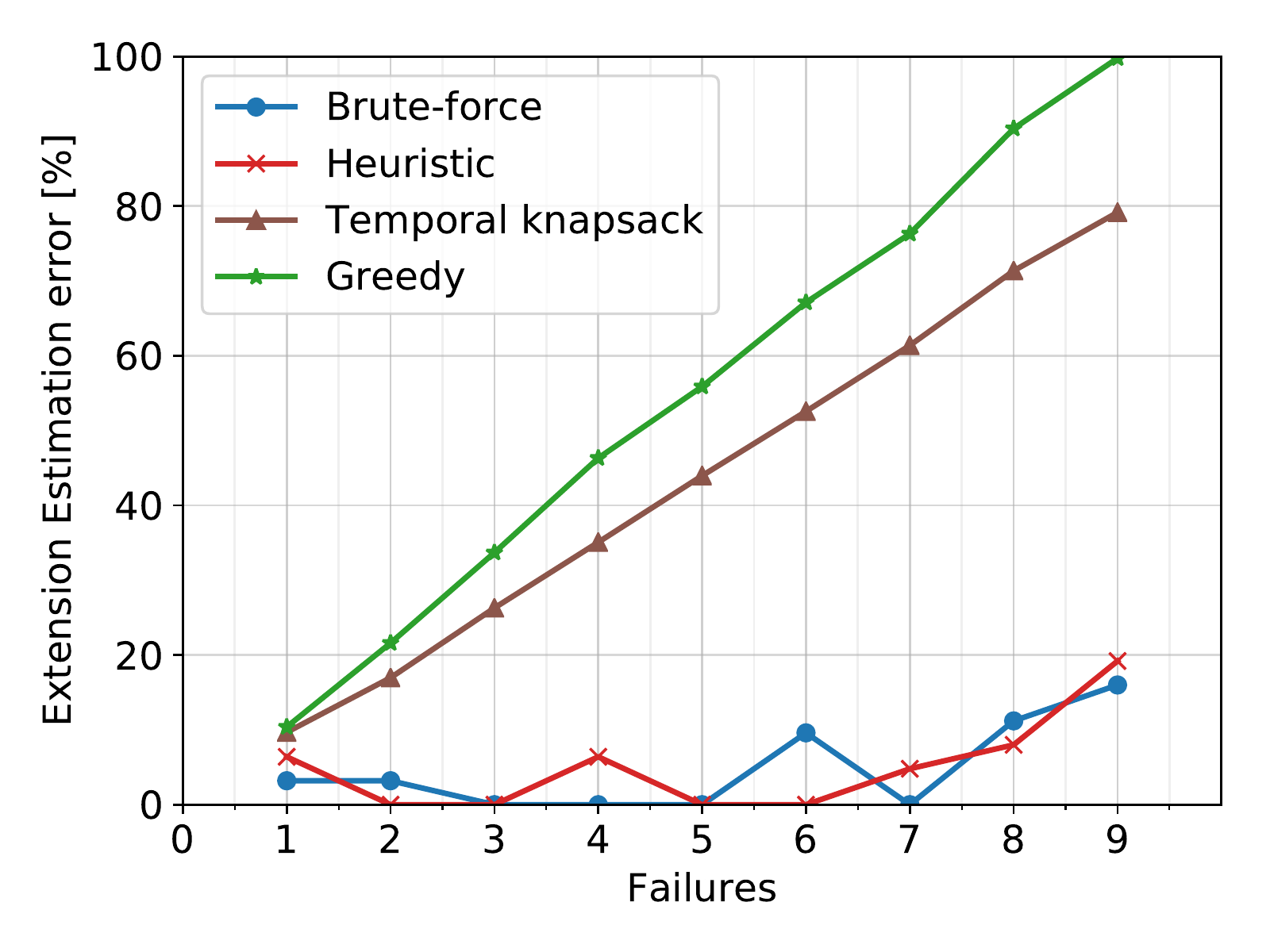}}
    \subfloat{\centering\includegraphics[width=.31\textwidth]{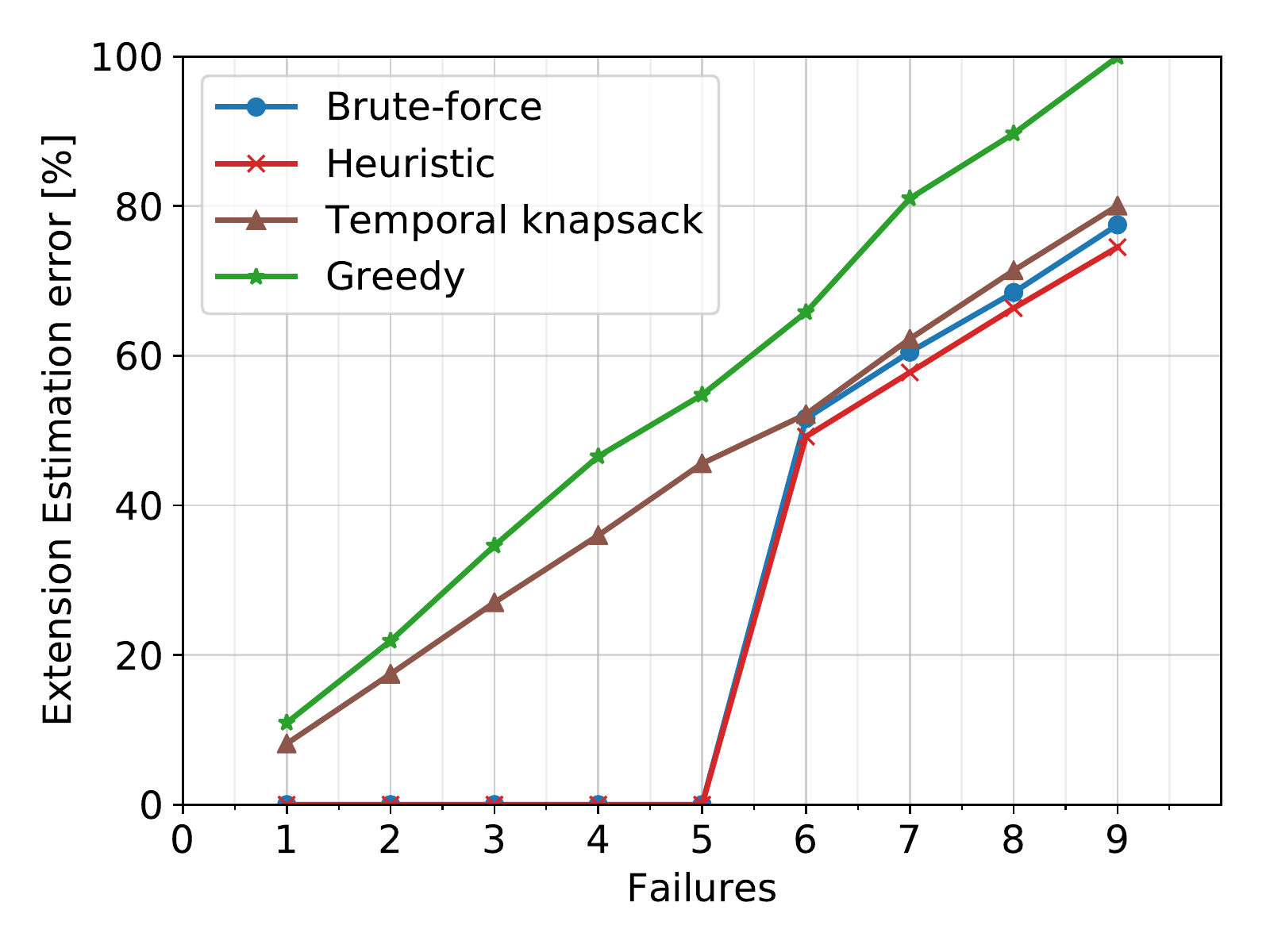}}
    \vspace{-0.2cm}
     \caption{\small Effectiveness of the elastic composition in a failure (a) Risk-taker users $w_r(Rel(Comp_i) \in [0.1,0.3]$ (b) Risk-averse users $w_r(Rel(Comp_i) \in [0.7,0.9]$ (c) Risk-neutral users$w_r(Rel(Comp_i) \in [0.4,0.6]$ }
\label{fig:effectivness}
    \vspace{-0.4cm}

\end{figure*}
{\em Efficiency}, in terms of the expected query time extension, refers to the difference between the optimal composite IoT energy services sets (i.e., Pareto front) obtained by the heuristic-based elastic composition and the brute force based composition. The Pareto front set contains the optimal set of composite IoT energy services in terms of the maximum reliability and the minimum extension time after the soft deadline of the query. First, we compare the number of candidate optimal compositions (i.e., cardinality) in each Pareto front set. Fig. \ref{fig:expected1} demonstrates the variations of cardinality of Pareto front sets for an energy query by increasing the number of services. The heuristic always provides a lower number of optimal composite IoT energy services. Second, we evaluate the efficiency of the heuristic by comparing the {\em expected time extension} obtained by the heuristic and the brute force based compositions for multiple energy queries. We use the reliability weight $w_r$ to ponder the utility function $u$ (see equation \ref{eq:utility}). We run both composition techniques for 5000 IoT energy queries for each reliability weight value ranging from $1$ to $9$. We consider the average expected time extension for each reliability weight (see Fig. \ref{fig:expected2}). {\em The heuristic estimates almost the same time extensions provided by the Brute-force}. The elastic composition expects (i) short time extension when the reliability weight is low, because the selected component services provide high energy amounts and low reliability scores. The elastic composition also estimates (ii) longer time extension when higher reliability score is considered for component services. This could be explained by the Pareto front elements i.e., best composite candidates. Most of them are either high in terms of provided energy or reliability. If reliability has a low weight, services providing high amounts of energy are privileged. Thus, the estimated composition does not expect long time extension to fulfill the require energy. However, the {\em actual} time extension might be longer than  expected time extension in a failure-prone crowdsourced environment. In the following, we analyze the expected extension time error in different scenarios.


%




\vspace{-0.2cm}
\subsection{Effectiveness}
\vspace{-0.1cm}

We investigate the {\em effectiveness} of the proposed elastic composition framework. We compare the {\em expected}  and the {\em actual} time extensions of the query duration in a failure-prone crowdsourced IoT environment. The actual time extension of a query duration is the required time to replace the failed component services in a composite IoT energy service.



For each confined area, a number of services with {\em low reliability} are randomly selected and revoked. The reliability of an IoT energy service is calculated based on the fluctuating behavior of its provider (see section \ref{profiling}). The expected energy and time extension of a composite IoT energy service are estimated based on the service {\em advertisement} and the {\em fluctuating} behavior of services providers in advance. The effective energy and the actual time extension are calculated after the composition. 

We consider an {\em extension estimation error} $EXER$ by an elastic composition $C$ if the expected extension time of an energy query $Q$, i.e., $ExtQ(C)$ is {\em before} the {\em hard deadline} $Q.Dl_h$  of the query $Q$ and the actual extension time of the query $Q$ i.e., $EffQ(C)$ is {\em after} the {\em hard deadline} $Q.Dl_h$  of that query $Q$: ($ExtQ(C)\leq Q.Dl_h$ and $EffQ(C) > Q.Dl_h$). We gradually vary the number of randomly generated failures (i.e., revoked services) in a scale between $[1-10]$. We calculate the ratio of the elastic compositions with extension estimation errors to the total number of energy queries (see Fig. \ref{fig:effectivness}).

The elastic composition considers the aggregate reliability of a composite service and estimates the time extension after the soft deadline (i.e., end of the query duration). The optimal composite service is selected from the Pareto front set based on users' preferences using the utility $u$ (see equation \ref{eq:utility}). We evaluate the estimation of the time extension with three different preference strategies. Risk-averse users prefer to select high reliability weights $w_r\in [7~,~9]$ more than energy. Risk neutral users compromise reliability and energy equally  $w_r\in [4~,~6]$. Risk-taker users are greedy, they decide mostly based on energy $w_r\in [1~,~3]$. The error estimation in time extension for all the algorithms in the three risk strategies is {\em relatively low when failures are not frequent}. Similarly, the $EXER$ ratio is higher when more failures appear. 

We observe a similar behavior of all composition algorithms for the risk-taker behavior (see Fig. \ref{fig:effectivness} (a)). Reliable compositions are not selected by the elastic composition when the reliability wight is very low in the case of Risk-taker strategy. However, the risk-averse users set a high reliability weight. Thus, reliable compositions are selected which minimizes the error estimation in time extension for the elastic composition algorithms (see Fig. \ref{fig:effectivness} (b)). Risk-taker strategy prefers composite services with high energy which might have a low reliability score. If a component service providing high energy amount fails, a longer time is required to recover the energy loss. In contrast, risk-averse users do not have a long time extension because their selected services are reliable and less prone to failures. In a risk-neutral strategy, the elastic composition have a low error estimation values for less failure-prone environments $failure \leq 5$ (see Fig. \ref{fig:effectivness} (c)). Overall, the $EXER$ ratio is almost equal for the brute force and the heuristic based elastic compositions for all types of IoT energy consumers (risk-averse, risk-takers, and risk-neutral). The elastic composition provides the most reliable composition according to IoT energy consumers risk strategies and estimates the time extension with high accuracy, compared to the brute force based composition. 




\vspace{-0.2cm}
\section{Related work}
\vspace{-0.1cm}
The background of our work comes from three different areas, i.e., \textit{energy harvesting, energy sharing}, and \textit{service selection and composition}. We describe the related work to our research in each of these domains. 

 \subsection{Energy harvesting and wireless transfer}
 Recently, research has been conducted to integrate harvesting energy into designing IoT objects (i.e., wearables) \cite{gorlatova2015movers}\cite{khalifa2017harke}. Body heat and movement provide a significant source of energy for wearables \cite{choi2017wearable}. The Kinetic Energy Harvesting (KEH) for IoT captures kinetic energy from wearables while exercising different daily activities such as walking, running, and reading \cite{gorlatova2015movers}. The advent of wireless charging makes the harvested energy from IoT devices more flexible and convenient to be easily shared. Energy sharing helps create self-sustained systems. Different techniques have been developed for the wireless charging in IoT and sensor networks \cite{lu2016wireless}. The most common techniques are magnetic inductive coupling, magnetic resonant coupling, and microwave radiation. These techniques are used in wireless sensor networks by deploying charger robots in the network to charge the low battery sensors \cite{na2018energy}. A new paradigm of uncoupled wireless charging based on radio waves has emerged \cite{bell2019systems}. The Energous Wattup applies radio waves to enable wireless energy sharing for IoT devices. 

\subsection{Energy sharing}
Wireless crowd charging has been introduced recently to provide IoT users with ubiquitous power access through crowdsourcing \cite{lakhdari2020Vision}\cite{dhungana2020peer}\cite{raptis2019online}. Dhungana et al. present a recent survey on peer-to-peer energy sharing in several applications including wireless sensor networks (WSN) and mobile social networks (MSN) \cite{dhungana2020peer}. Raptis et al. claim that knowing the properties of the crowd social network is crucial in the design of crowd charging protocols. One main aspect of the crowd is the users' active presence in online social networks. They suggest the exploitation of online social information in order to tune the wireless crowd charging process \cite{raptis2019online}. In our paper, we use the service paradigm to abstract wireless energy services and facilitate energy sharing in the crowdsourced IoT environment.

\subsection{Service selection and composition}
Service composition and selection approaches are crucial for service computing and hence have got significant growing research work. Several research studies, e.g., \cite{tan2010data}\cite{wang2014constraint}, focused on composing services based on functional requirements. For example, Tan et al.  \cite{tan2010data} proposed a data-driven composition approach that uses Petri-nets to meet the application’s functional requirements.  Wang et al.  \cite{wang2014constraint} address the problem of service functionalities constraints by introducing a pre-processing technique and a graph search-based algorithm to compose services. Due to its significance, many recent studies including  \cite{deng2016service} have proposed service composition based on QoS properties. A service pruning method to address the QoS-correlation problem in service selection and composition  \cite{deng2016service}. Many other QoS-aware service selection and composition, e.g., \cite{Niu2017} apply different methods such as estimation of distribution algorithm based on Restricted Boltzmann Machine (rEDA) and non-deterministic multi-objective evolutionary algorithm to derive multi-objective optimal service composition.
The service selection and composition have been applied to emerging fields such as cloud computing, sensor-cloud services, and social networks \cite{aamir2017social}. In sensor-cloud, services are composed according to their spatio-temporal features. They also must fulfill consumer preferences (QoS). 
These contributions are among the early attempts to compose crowdsourced energy services in the IoT environment.
\vspace{-0.25cm}
\section{Conclusion}
\vspace{-0.1cm}
We propose a novel elastic composition framework to crowdsource wireless energy services from IoT devices. We formulate the composition problem as a multi-objective optimization to meet users' energy requirements and to maximize the composition reliability with the minimum time extension after the soft deadline. We conduct a set of experiments on  real datasets to investigate the scalability and the effectiveness of the proposed approach. 
Results show that the proposed elastic composition is able to select the most reliable composition with the minimal time extension. In future work, we will explore the intermmittent behavior of energy services. i.e., both service providers and consumers are highly dynamic in both confined and open places.

\bibliographystyle{IEEEtran}
\bibliography{ICWS2019}

\end{document}